\documentclass[%
reprint,
superscriptaddress,
amsmath,amssymb,amsthm,mathrsfs,
aps,
prl,
]{revtex4-1}
\usepackage{graphicx}% Include figure files
\usepackage{dcolumn}% Align table columns on decimal point
\usepackage{bm}% bold math
\usepackage[usenames]{color}
\usepackage[ %showframe,%Uncomment any one of the following lines to test
margin=0.75in,
]{geometry}

\def\NCCF{NaCaCo$_2$F$_7$}
\def\NSCF{NaSrCo$_2$F$_7$}
\def\Tf{$T_f$}
\def\muB{$\mu_\mathrm{B}$}

\begin{document}

\title{
Real-space investigation of short-range magnetic correlations in fluoride pyrochlores NaCaCo$_2$F$_7$ and NaSrCo$_2$F$_7$ with magnetic pair distribution function analysis
}

\author{Benjamin A. Frandsen}
\email{benfrandsen@berkeley.edu}
\affiliation{%
	Materials Sciences Division, Lawrence Berkeley National Laboratory, Berkeley, California 94720, USA.
}%
	\affiliation{ %
	Department of Physics, University of California, Berkeley, California 94720, USA.
} %

\author{Kate A. Ross}
\affiliation{
Department of Physics, Colorado State University, Fort Collins, Colorado, 80523, USA.
}
\affiliation{
Quantum Materials Program, Canadian Institute for Advanced Research, Toronto, Ontario M5G 1Z8, Canada.
}

\author{Jason W. Krizan}
\affiliation{
Department of Physics, Princeton University, Princeton, New Jersey 08544, USA.
}

\author{G{{\o}}ran J. Nilsen}
\affiliation{
Institut Laue-Langevin, CS 20156, 38042 Grenoble C\'edex 9, France.
}
\affiliation{
	ISIS Neutron and Muon Facility, Rutherford Appleton Laboratory, Didcot OX11 0QX, United Kingdom.
}

\author{Andrew R. Wildes}
\affiliation{
	Institut Laue-Langevin, Grenoble 38000, France.
}

\author{Robert J. Cava}
\affiliation{
	Department of Physics, Princeton University, Princeton, New Jersey 08544, USA.
}

\author{Robert J. Birgeneau}
\affiliation{ %
	Department of Physics, University of California, Berkeley, California 94720, USA.
} %
\affiliation{%
	Materials Science Division, Lawrence Berkeley National Laboratory, Berkeley, California 94720, USA.
}%
\affiliation{ %
	Department of Materials Science and Engineering, University of California, Berkeley, California 94720, USA.
} %

\author{Simon J. L. Billinge}
\affiliation{%
	Condensed Matter Physics and Materials Science Department, Brookhaven National Laboratory, Upton, NY 11973, USA
}%
\affiliation{Department of Applied Physics and Applied Mathematics, Columbia University, New York, NY 10027, USA}

\begin{abstract}
We present time-of-flight neutron total scattering and polarized neutron scattering measurements of the magnetically frustrated compounds NaCaCo$_2$F$_7$ and NaSrCo$_2$F$_7$, which belong to a class of recently discovered pyrochlore compounds based on transition metals and fluorine. The magnetic pair distribution function (mPDF) technique is used to analyze and model the total scattering data in real space. We find that a previously-proposed model of short-range XY-like correlations with a length scale of 10-15~\AA, combined with nearest-neighbor collinear antiferromagnetic correlations, accurately describes the mPDF data at low temperature, confirming the magnetic ground state in these materials. This model is further verified by the polarized neutron scattering data. From an analysis of the temperature dependence of the mPDF and polarized neutron scattering data, we find that short-range correlations persist on the nearest-neighbor length scale up to 200~K, approximately two orders of magnitude higher than the spin freezing temperatures of these compounds. These results highlight the opportunity presented by these new pyrochlore compounds to study the effects of geometric frustration at relatively high temperatures, while also advancing the mPDF technique and providing a novel opportunity to investigate a genuinely short-range-ordered magnetic ground state directly in real space.
\end{abstract}

\maketitle

\section{Introduction}

The pyrochlore network of corner-sharing tetrahedra provides a versatile platform for studying the novel properties of geometrically frustrated magnetic systems. In these systems, the triangular motifs of the lattice prevent the magnetic interactions from being fully satisfied, often leading to macroscopic ground-state degeneracy and exotic collective excitations~\cite{balen;n10}. The frustrated magnetic interactions may include exchange interactions, anisotropy energies, dipolar interactions, the Dzyaloshinskii-Moriya interaction, and others. If these competing interactions are sufficiently balanced, thermal or quantum fluctuations can prevent long-range magnetic order from forming down to exceedingly low or even zero temperature, resulting in disordered ground states. Even systems that develop long-range order at sufficiently low temperature may still exhibit unconventional behavior, such as unusual magnetic symmetries or exotic mechanisms for the phase transition~\cite{gardn;rmp10}. The rare-earth titanate pyrochlores with the chemical formula $R_2$Ti$_2$O$_7$ ($R$ = rare earth) can be taken as exemplars of the rich physics resulting from geometric frustration: Spin ices with magnetic monopole excitations, candidate quantum spin liquids with fractional particle excitations, topological order, and phases selected by order-by-disorder have all been identified among the rare-earth titanate pyrochlores, stimulating vigorous experimental and theoretical investigation~\cite{gardn;rmp10,henle;arocmp10,caste;arocmp12}.

% a few sentences explaining why the pyrochlore structure can lead to a disordered magnetic ground state including what assumptions one needs to make about the magnetic interactions.  You should also explain what states are possible including psi1, psi2 and psi3.  Are there others and if not, why not?

Recently, a newly discovered family of transition-metal magnetic pyrochlores based on fluorine rather than oxygen has attracted significant interest~\cite{kriza;prb14,kriza;jpcm15,kriza;prb15,ross;prb16,sarka;arxiv16,sande;jpcm17,ross;prb17}. These materials open up new avenues of research into frustrated magnetism, particularly given the novel addition of magnetic 3$d$ transition-metal ions into the structure. Unlike the rare-earth pyrochlores, in which the relatively weak interactions among the 4$f$ electrons generally restrict the most interesting phenomena to millikelvin-scale temperatures, the larger interaction strengths of the 3$d$ electrons in these new materials allow frustration-related physics to be accessed at kelvin scales.

\begin{figure}
	\includegraphics[width=80mm]{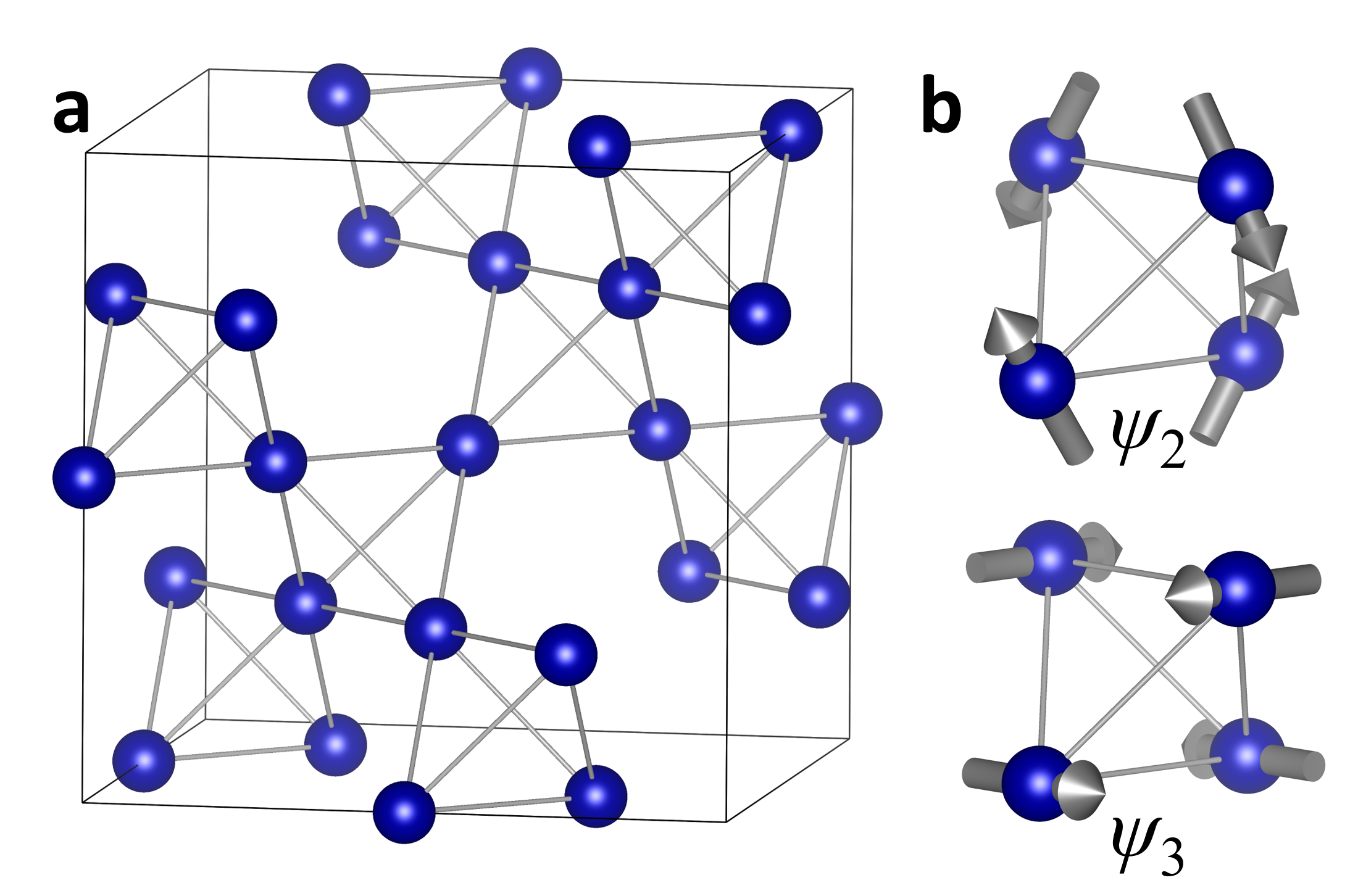}
	\caption{\label{fig:structure} (Color online)  (a) Pyrochlore network of corner-sharing tetrahedra formed by the Co atoms in \NCCF\ and \NSCF. The Na and Ca/Sr atoms (not shown) also form a pyrochlore network. (b) Magnetic moment configurations for the $\psi_2$ and $\psi_3$ basis states of the $\Gamma_5$ irreducible representation of the tetrahedral point group.}
	
\end{figure}

The first of these new fluoride pyrochlores to be synthesized as large single crystals was \NCCF, followed by the similar compound \NSCF~\cite{kriza;prb14,kriza;jpcm15}. In these materials, the cubic structure (space group $Fd\bar{3}m$) hosts two pyrochlore sublattices, one with the Co$^{2+}$ ions and the other with the randomly distributed Na$^+$ and Ca$^{2+}$/Sr$^{2+}$ ions. The Co sublattice is shown in Fig.~\ref{fig:structure}(a). Magnetic susceptibility measurements of \NCCF\ revealed large moments of $\sim$6.1~\muB\ per Co$^{2+}$, a Curie-Weiss temperature of $\Theta_{\mathrm{CW}}=-140$~K, and a spin freezing temperature of \Tf~=~2.4~K, indicating a high level of magnetic frustration ($f=|\Theta_{\mathrm{CW}}|/T_f=58$)~\cite{kriza;prb14}. \NSCF\ has nearly identical properties, with a moment size of 5.9~\muB, $\Theta_{\mathrm{CW}}=-127$~K, and \Tf~=~3~K~\cite{kriza;jpcm15}.

%%%%%%%%%%%%%%%%%
% Begin Figure
%%%%%%%%%%%%%%%%%
%%%%%%%%%%%%
% End Figure
%%%%%%%%%%%%

Elastic and inelastic neutron scattering measurements have been carried out previously on single crystals of \NCCF\ and \NSCF\ to better understand the nature of the magnetic ground state~\cite{ross;prb16,ross;prb17}. Below \Tf, the elastic scattering takes the form of diffuse spots, indicating a genuinely short-range-ordered (SRO) but highly correlated magnetic ground state. On the basis of the locations of these diffuse spots as well as systematic absences of intensity measured in the $(HHL)$ scattering plane, the proposed magnetic model consists of SRO XY-type antiferromagnetic (AF) clusters with correlation lengths of 16.7~\AA\ (\NCCF) and 17.6~\AA\ (\NSCF). These clusters are built from linear combinations of the form $\cos (\alpha) \psi_2 + \sin (\alpha) \psi_3$, where $\psi_2$ and $\psi_3$ are the basis vectors of the $\Gamma_5$ irreducible representation (irrep) of the tetrahedral point group for $Q = 0$ wave vectors. This irrep is two dimensional, so $\psi_2$ and $\psi_3$ are the only basis vectors. They are illustrated in Fig.~\ref{fig:structure}(b). The inelastic magnetic scattering below \Tf\ consists of a low-energy ($\Delta E \lesssim 2$~meV) signal with a pattern similar to that of the elastic scattering, indicating fluctuations within the $\Gamma_5$ manifold, and a higher-energy ($2 \lesssim \Delta E \lesssim 8$~meV) signal consistent with fluctuating collinear AF correlations with a much shorter correlation length. These distinct inelastic signals were observed to persist up to at least 14~K (the highest temperature measured), whereas the elastic signal vanished above \Tf.

Finally, the single-ion levels of Co$^{2+}$ were investigated by inelastic neutron scattering over much larger energy transfers (up to 1500~meV)~\cite{ross;prb17}. It was found that the crystal field splits the $d$ orbitals in such a way that the lowest-energy orbital is separated from the next closest level by 30~meV, corresponding to approximately 300~K. At low temperature, therefore, the higher energy states are frozen out. By Kramer's Theorem, the ground state is doubly degenerate, resulting in a two-state system. Hence, the system can be described with $J_{\mathrm{eff}}=1/2$, even though the magnitude of the full single-ion moment is approximately 6~$\mu_{\mathrm{B}}$ for both compounds. This large moment indicates that both spin and orbital angular momentum contribute to the magnetism. Consequently, the single-ion magnetic moment may display an anisotropic Zeeman splitting in an applied magnetic field. The rank-two tensor that encodes this anisotropic response, called the $g$ tensor, was shown in earlier work to cause a significantly enhanced Zeeman response in the plane perpendicular to the local [111] direction for each ion, and a much smaller response out of the plane. In such a situation, the single-ion moments and $g$ tensor are referred to as XY-like.

Although the previous work provides an excellent starting point for understanding the frustrated magnetism in \NCCF\ and \NSCF, it would be valuable to confirm the proposed XY model through neutron scattering that probes a larger region of reciprocal space than the single scattering plane examined thus far. Measurements at higher temperatures that approach or exceed $|\Theta_{\mathrm{CW}}| \approx 140$~K would likewise be highly informative. In this work, we present magnetic pair distribution function (mPDF) analysis~\cite{frand;aca14,frand;aca15} of neutron total scattering measurements up to 50~K and polarized neutron diffraction measurements up to 300~K, both using powder samples of \NCCF\ and \NSCF. The results confirm the model of SRO XY-type correlations proposed previously and reveal the presence of magnetic correlations up to 200~K. This work also represents the first application of the recently developed mPDF technique to a geometrically frustrated material with a short-range-ordered magnetic ground state, offering a rare opportunity to observe genuinely short-range magnetic correlations directly in real space.

\section{Experimental Details}

Single-crystal samples of \NCCF\ and \NSCF\ were grown by the floating zone method as described in Refs.~\onlinecite{kriza;prb14,kriza;jpcm15} and then ground into a fine powder in a mortar and pestle. Time-of-flight neutron total scattering measurements of both compounds were performed on the NOMAD instrument at the Spallation Neutron Source (SNS) of Oak Ridge National Laboratory (ORNL)~\cite{neufe;nimb12}. Polarized neutron diffraction measurements of \NCCF\ were performed on the D7 instrument at the Institut Laue Langevin (ILL)~\cite{stewa;jac09}, with the $xyz$ polarization method~\cite{schar;pssa93} used to separate the magnetic, nuclear, and nuclear spin incoherent scattering cross sections. An incident wavelength of 3.12~\AA\ was used to obtain the maximum momentum transfer possible on the D7 instrument ($Q_{\mathrm{max}} = 3.91$~\AA$^{-1}$). At both beamlines, the temperature was controlled using a liquid helium cryostat.

The total scattering data collected on NOMAD were reduced and Fourier transformed to the real-space pair distribution function (PDF) using the automatic data reduction scripts at the beamline. Any magnetic scattering present in the data is included in the Fourier transform, and as such, the transformed data contain both the atomic PDF and the mPDF. The mPDF provides an intuitive description of the local magnetic structure directly in real space. Since it is generated from the Fourier transform of the total scattering, including both Bragg and diffuse scattering, it is sensitive to long-range and short-range magnetic correlations. Roughly speaking, a positive peak in the mPDF at given value of real-space distance $r$ corresponds to ferromagnetic correlations between spins separated by that distance, and a negative peak corresponds to antiferromagnetic correlations. A detailed derivation and description of the mPDF is provided in Ref.~\onlinecite{frand;aca14}. When the mPDF is obtained by Fourier transforming the magnetic scattering without normalizing by the magnetic form factor, as is the case for the present work, the resulting quantity is broadened out in real space by a factor of approximately $\sqrt{2}$ times the real-space extent of the individual local magnetic moments. We will nevertheless continue to refer to this as the mPDF in the present work. This is described in more detail in Ref.~\onlinecite{frand;aca15}. Because the focus here is on the magnetic rather than atomic structure, a conservative $Q_{\mathrm{max}}$ of 20~\AA$^{-1}$ was used for the Fourier transform. Modeling of the atomic PDF was carried out with the PDFgui program~\cite{farro;jpcm07}, and the mPDF was modeled with the open-source package diffpy.mpdf, part of the DiffPy suite of programs for diffraction analysis~\cite{juhas;aca15}.

\section{Results and Discussion}

\begin{figure}
	\includegraphics[width=80mm]{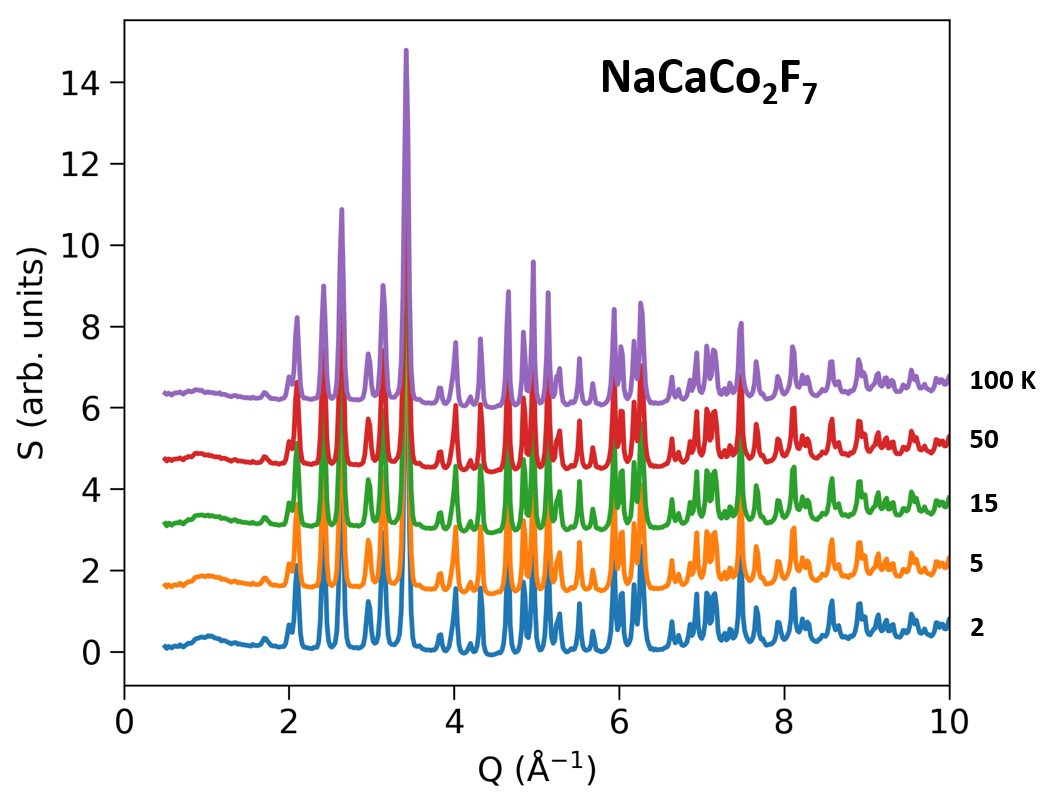}
	\caption{\label{fig:SQ-PDF} (Color online). Total scattering structure function $S(Q)$ for \NCCF\ at various temperatures collected on the NOMAD beamline, offset vertically for clarity. The diffuse feature centered around 1~\AA$^{-1}$ arises from magnetic scattering and gradually weakens with increasing temperature.}
	
\end{figure}

\textit{Total scattering measurements and extraction of the mPDF.} We first present the results of the neutron total scattering experiments performed on the NOMAD instrument and our subsequent procedure to extract the mPDF from the data. \NCCF\ was measured at 2~K, 5~K, 15~K, 50~K, and 100~K. The structure function $S(Q)$ is displayed up to 10~\AA$^{-1}$ for each of these temperatures in Fig.~\ref{fig:SQ-PDF}. The magnetic contribution to the structure function is most visible for the data collected at 2~K, where a broad hump of diffuse scattering centered around 1~\AA$^{-1}$ is visible, matching the position of the diffuse spots observed in the earlier work on a single-crystal sample. This feature becomes progressively weaker as the temperature is raised, although it is still observable even at 100~K. A more thorough temperature dependence will be presented later with the polarized neutron diffraction results. Similar total scattering data were obtained for \NSCF\ at 2~K, 15~K, 50~K, and 100~K.

%%%%%%%%%%%%%%%%%
% Begin Figure
%%%%%%%%%%%%%%%%%
%%%%%%%%%%%%
% End Figure
%%%%%%%%%%%%

\begin{figure}
	\includegraphics[width=80mm]{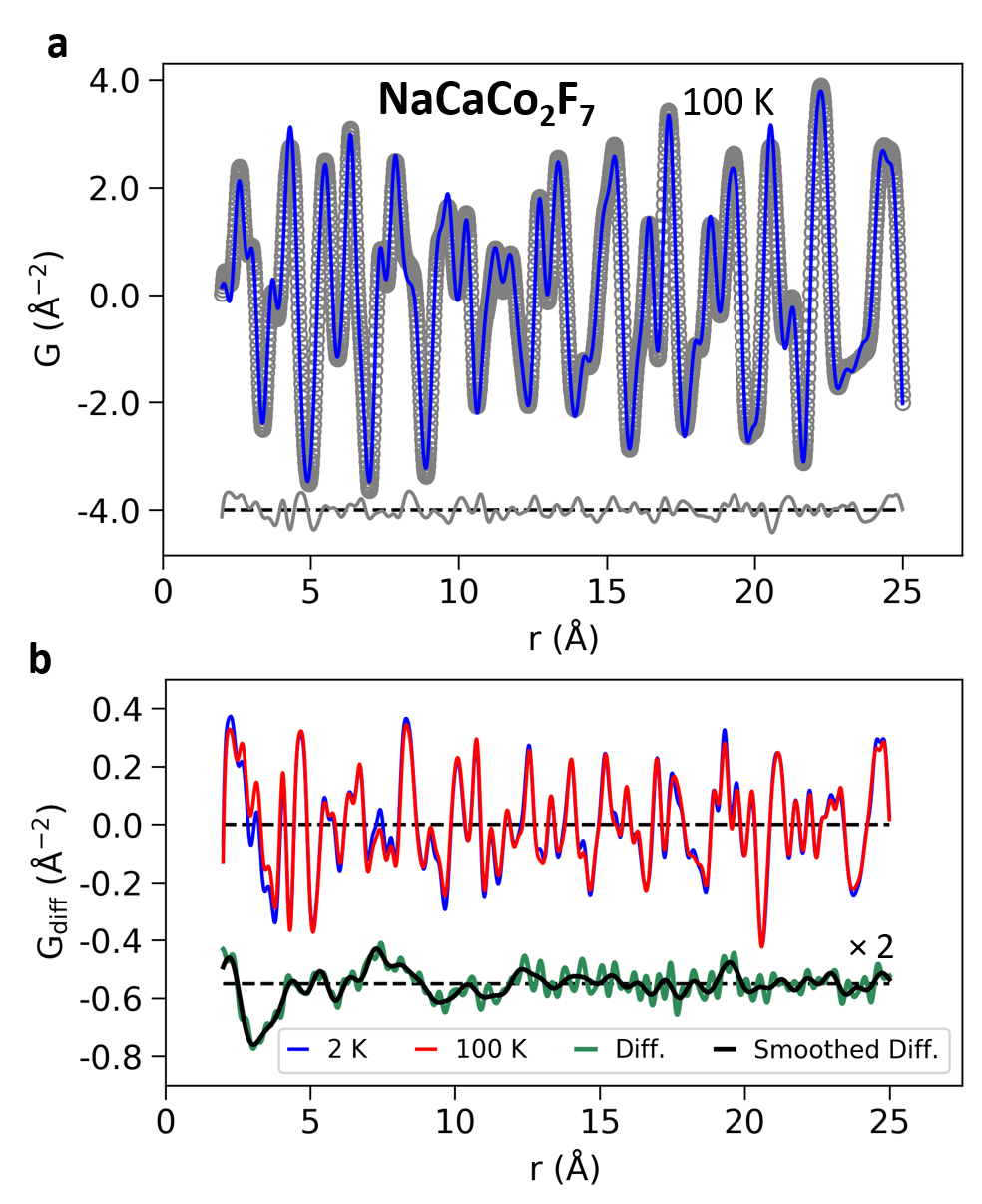}
	\caption{\label{fig:PDF-diffs} (Color online). (a) Atomic PDF refinement for \NCCF\ at 100~K. The gray circles represent the data, the blue curve the refined PDF, and the lower gray curve the fit residual, offset for clarity. (b) Fit residuals for the atomic PDF refinements at 100~K (red) and 2~K (blue), with the difference between the two fit residuals shown as the lower green curve, multiplied by two and offset vertically for clarity. The lower black curve is the difference in fit residuals after filtering out high-frequency components (see main text). This difference curve contains the mPDF at 2~K.}
	
\end{figure}

The PDF generated from the 100~K data for \NCCF\ is shown as gray circles in Fig.~\ref{fig:PDF-diffs}(a), with the calculated PDF from the refined cubic model shown as the blue line and the fit residual shown in gray below. The model describes the data fairly well, validating the quality of the sample.

The relative weight of the mPDF to the atomic PDF in the total PDF signal is given by approximately
\begin{align}
	\frac{N_s}{N_a}\frac{\frac{2}{3}(\frac{\gamma r_0}{2})^2(gJ)^2}{2\pi\langle b \rangle^2},\label{eq;scaleFactor}
\end{align}
where $N_s$ and $N_a$ are the number of spins and atoms per unit cell, respectively, $\gamma = 1.913$ is the neutron magnetic moment in units of nuclear magnetons, $r_0=\frac{\mu _0}{4\pi}\frac{e^2}{m_e}$ is the classical electron radius, the product $gJ$ gives the locally ordered magnetic moment in units of Bohr magnetons, and $\langle b \rangle$ is the average nuclear scattering length of the material. For \NCCF\ and \NSCF\, this ratio is approximately 0.15, assuming fully ordered correlations between neighboring magnetic moments. Additionally, the mPDF peaks are broadened by approximately $\sqrt{2}$ times the real-space width of the wavefunctions of the electrons forming the local magnetic moments, causing the peak heights of the mPDF to be further reduced by a factor of $\sim$4 in \NCCF\ and \NSCF. Altogether then, the average magnitude of the mPDF signal in these materials is expected to be less than 5\% that of the atomic PDF.

Given the relatively small strength of the mPDF signal compared to that of the atomic PDF, care must be taken when attempting to extract the mPDF from the overall PDF data. As has been done in previous studies, we first refined the atomic structure and subtracted the best-fit atomic PDF from the data at each temperature. The resulting fit residuals contain the mPDF as well as experimental errors, such as imperfect background subtractions or absorption corrections, and any errors remaining from the structural model. In some cases, such as for MnO, the mPDF is much larger than these other contributions, so the fit residual after subtracting the refined atomic PDF can be used directly for mPDF analysis. However, this is not the case for \NCCF\ or \NSCF; the errors are comparable in magnitude to the mPDF signal we wish to observe. Additional steps are therefore required to accurately extract the mPDF.

Recognizing that the fit residuals are dominated by features that are independent of temperature and therefore unrelated to the mPDF, we subtracted the 100-K fit residual from that of each lower temperature. This removed the temperature-independent portion of the fit residuals, leaving the temperature-dependent mPDF. We note that this is somewhat problematic because a small amount of magnetic scattering is still present at 100~K, but we expect this to have a minimal impact on the mPDF data, and it gives us the benefit of removing the larger temperature-independent background that would otherwise obscure the data. The result of this procedure for 2~K is illustrated in Fig.~\ref{fig:PDF-diffs}(b), with the atomic fit residual at 100~K and 2~K shown as the red and blue curves, respectively, and the difference between the two shown as the green curve, offset vertically and multiplied by two for clarity. Then, noting that the unnormalized mPDF is dominated by scattering at low $Q$ due to the magnetic form factor, we performed a Fourier filter operation on the difference of fit residuals[i.e., the green curve in Fig.~\ref{fig:PDF-diffs}(b)] to remove the high-frequency components. As the cutoff $Q$ value for the Fourier filter, we chose the point where the magnetic form factor is reduced to 1\% of its maximal value, which occurs at 8~\AA$^{-1}$ in these materials. The resulting smoothed signal is shown as the black curve in Fig.~\ref{fig:PDF-diffs}(b).

% by Fourier transforming the residual back into momentum space, truncating the data beyond some cutoff value of $Q$, and transforming into real space once more.... A similar approach is commonly used in standard neutron diffraction to isolate magnetic scattering. 
%%%%%%%%%%%%%%%%%
% Begin Figure
%%%%%%%%%%%%%%%%%
%%%%%%%%%%%%
% End Figure
%%%%%%%%%%%%

Since the mPDF is quite weak relative to the overall scale of the total PDF, it is important to carefully evaluate its reliability. Multiple considerations reassure us that the mPDF obtained this way is accurate. First, it has a structure that is consistent with the expectations for short-range, primarily antiferromagnetic correlations: the peaks are centered around $r$ values corresponding to Co-Co distances, they alternate between negative and positive values due to the antiferromagnetic alignment, and their magnitude diminishes with increasing $r$ due to the finite correlation length. Additionally, the mPDF signal shows a reasonable temperature dependence, gradually decreasing as the temperature rises. Finally, and most importantly, the mPDF data can be modeled to a high degree of accuracy using the SRO magnetic structure reported previously. The success of these fits, which we describe below, demonstrates that the data obtained through this procedure are reliable and that the proposed magnetic model is accurate.

\textit{Modeling the temperature-dependent mPDF.} The mPDF data can be successfully modeled in a straightforward manner. Starting with the refined structural parameters from the atomic PDF fit, we first calculated the mPDF from a configuration in which every tetrahedron has the same arrangement of moments selected from the linear combination $\cos (\alpha) \psi_2 + \sin (\alpha) \psi_3$, with some arbitrarily chosen value of $\alpha$. The mPDF is completely independent of $\alpha$, so the particular value used does not matter. To simulate the short-range nature of the magnetic correlations, we then applied an exponential damping function of the form $\exp(-r/\xi)$, where $\xi$ represents the correlation length. The best fit was then obtained by performing a least-squares refinement of $\xi$ and an overall scale factor.

\begin{figure}
	\includegraphics[width=60mm]{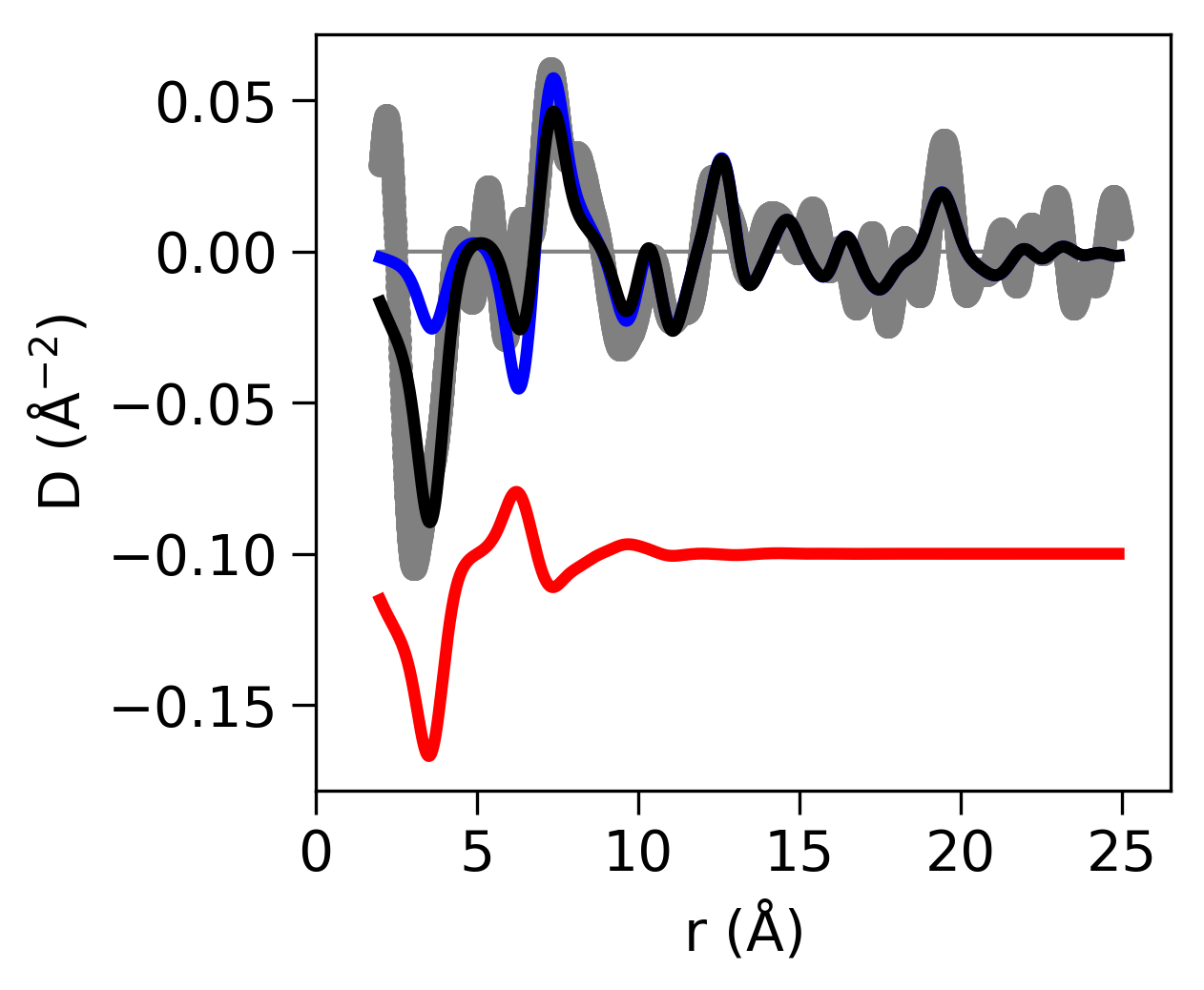}
	\caption{\label{fig:one} (Color online) Magnetic PDF refinement for \NCCF\ at 2~K. The gray curve represents the data, the blue curve is the best fit using a single-component model containing short-range XY correlations, the black curve shows the result when a second component with collinear antiferromagnetic correlations is included, and the red curve shows the contribution from this second component, offset vertically for clarity.}
	
\end{figure}

The result of this procedure for \NCCF\ at 2~K is shown in Fig.~\ref{fig:one}, in which the gray curve is the experimental mPDF, labeled $D(r)$, and the blue curve the simulated mPDF. The agreement is quite good for $r \gtrsim 7$~\AA, indicating that this magnetic structure explains well the magnetic correlations on that length scale. However, the model provides a very poor fit in the low-$r$ region, specifically the first two Co-Co neighbor distances of $\sim$3.7~\AA\ and $\sim$6.4~\AA, which suggests that XY correlations alone cannot fully explain the observed magnetic correlations on this very short length scale. We also tried a model in which the decay of the magnetic correlations is Gaussian rather than exponential, but this did not improve the fit. Instead, a good fit can be obtained by including as an additional component the mPDF generated from very short-range collinear AF correlations, with an independently refined scale factor and correlation length. The result of this two-component fit, shown as the black curve in Fig.~\ref{fig:one}, nicely captures the observed mPDF across the full range of $r$. We note that the small oscillations not captured by the model can be ruled out as genuine mPDF features because they are either not centered around Co-Co distances and/or are too narrow given that the magnetic scattering was not normalized by the magnetic form factor before transforming to the mPDF. The collinear AF component, shown as the red curve offset vertically downward in Fig.~\ref{fig:one}, has a refined scale factor approximately twice that of the XY component and is strongly damped for correlations beyond the second nearest neighbor distance.
%%%%%%%%%%%%%%%%%
% Begin Figure
%%%%%%%%%%%%%%%%%
%%%%%%%%%%%%
% End Figure
%%%%%%%%%%%%

This second AF component can be naturally explained on the basis of the earlier neutron scattering work on a single-crystal specimen. As mentioned previously, two distinct patterns were observed in the inelastic channel at 1.7~K: a low-energy signal with $\Delta E \lesssim 2$~meV arising from fluctuations within the $\Gamma_5$ irrep, and a slightly higher-energy signal with $2 \lesssim \Delta E \lesssim 8$~meV due to shorter-range fluctuating collinear AF correlations. Since the NOMAD instrument does not employ any energy analysis, the PDF data contain information from both elastic and inelastic scattering, yielding the equal-time correlation function from static and dynamic correlations within an instrument-specific effective energy window. For NOMAD, this energy window is sufficiently large to integrate over both of these bands of magnetic fluctuations.

The slow fluctuations within the $\Gamma_5$ manifold would be indistinguishable from the static clusters, aside from possible differences in the correlation length. We therefore expect that both the static and slowly fluctuating $\Gamma_5$ correlations are captured by the mPDF pattern calculated from the simulated XY clusters described above. However, we attribute the additional mPDF component at the nearest-neighbor distance to the higher-energy collinear AF fluctuations. The difference in scale factors of the two components is consistent with the ratio of the spectral weight of the elastic and inelastic contributions to the scattering pattern. Inspection of the fit shown in Fig.~\ref{fig:one} shows that this collinear AF component is most prominent for the first two Co-Co distances, indicating that the correlations persist appreciably only over this short length scale. This agrees with the suggestion from Ref.~\onlinecite{ross;prb16} that the correlation length of the dynamic collinear AF fluctuations extends across only one or two tetrahedra. We note that the collinear AF model used here, which exhibits a reversal of magnetization between two of the four tetrahedra in the unit cell, differs from the model used in the analysis of the single-crystal, for which all tetrahedra are equivalent. Both models produce an identical mPDF peak at the nearest neighbor distance, correcting the main deficiency of the single-component model, while the model with alternating magnetization provides a slightly better fit at the second-nearest neighbor distance. However, the short correlation length makes it difficult to definitively confirm one collinear AF model over the other.

The refined correlation length of the XY clusters determined from the two-component fit is 8.4~$\pm$~0.5~\AA. Significant features in the mPDF are visible out to larger $r$, such as the peak around 19~\AA, but these are actually much smaller in magnitude than they would be for the undamped mPDF pattern. The refined correlation length for the collinear AF component is 1.9~$\pm$~0.1~\AA. The XY correlation length we find here is in qualitative but not quantitative agreement with that determined from the width of the elastic peak measured from a single-crystal sample at 1.7~K, reported to be $\xi = 16 \pm 1$~\AA. We attribute the discrepancy between these results primarily to the fact that the mPDF data contain not only the elastic scattering from static XY correlations, but also the low-energy inelastic scattering from shorter-range dynamical XY correlations, resulting in a shorter average correlation length than would be expected from the purely elastic signal. This is in contrast to the single-crystal study, which reported the correlation length based solely on the elastic scattering. We further note that the fit can be improved somewhat by including a third component consisting solely of magnetic correlations within [111]-type sheets, providing tentative confirmation of the anisotropic correlation length suggested by the shape of the diffuse spots in the single-crystal measurements. However, this additional component causes increased correlation between the refined parameters, so we will use the more robust two-component model for the remaining analysis. %In addition, the presence of both the XY and collinear AF contributions in the mPDF creates some challenges for refining reliable correlation lengths, since the magnitude of one component will have a significant impact on the refined magnitude of the other component, and hence its correlation length. For example, if we include the contribution from the collinear AF component only up to the nearest neighbor, the refined XY correlation length increases to 11.4~$\pm$~0.8~\AA.

%%%%%%%%%%%%%%%%%
% Begin Figure
%%%%%%%%%%%%%%%%%
\begin{figure}
	\includegraphics[width=80mm]{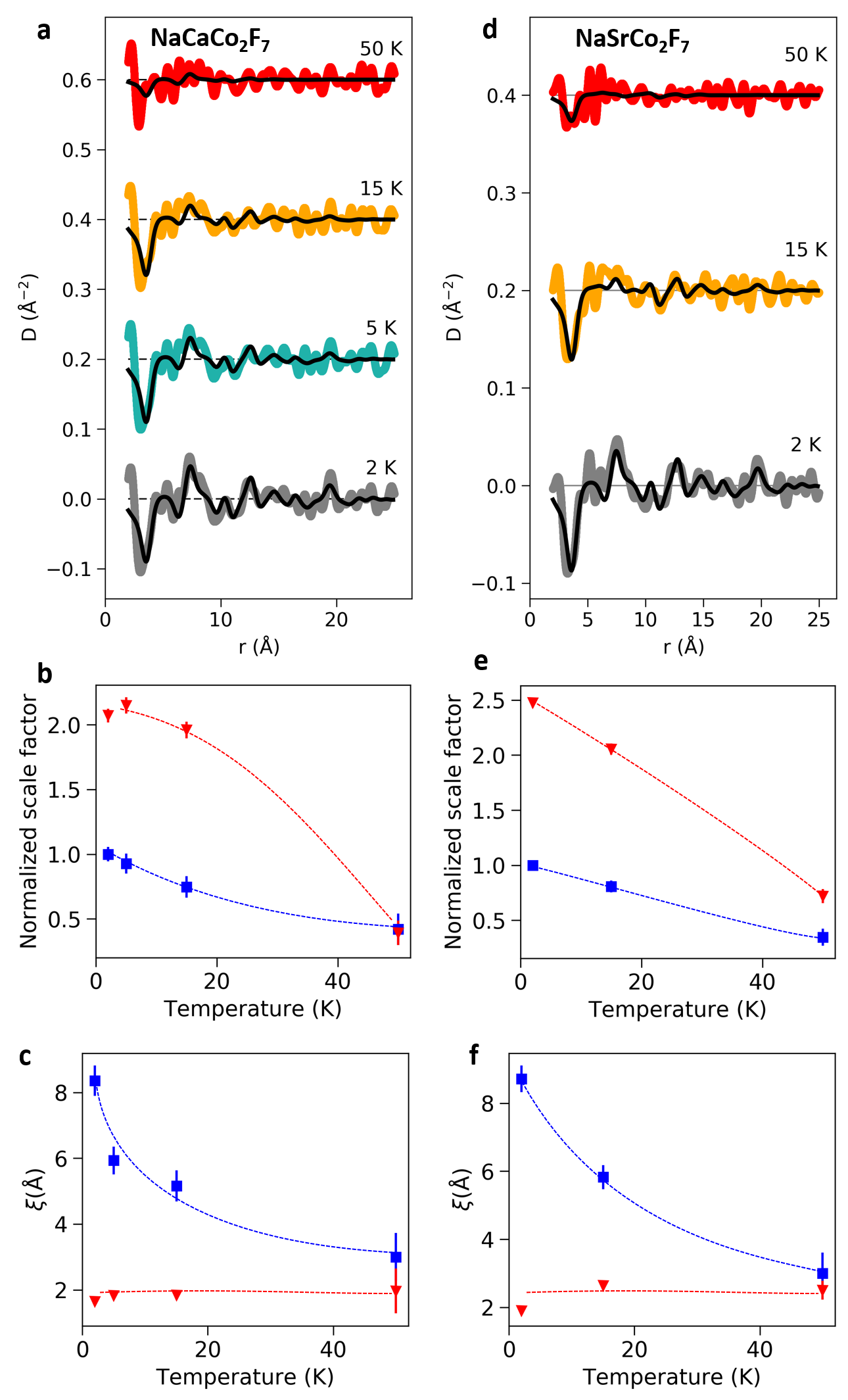}
	\caption{\label{fig:both} (Color online) (a) Two-component mPDF fits for \NCCF\ at various temperatures, offset vertically for clarity. The colored curves represent the data, the black curves the fits. (b) Refined scale factors for the fits in (a), normalized by the maximal value of the XY component. The blue squares represent the scale factor for the XY component, the red triangles for the collinear antiferromagnetic component. (c) Refined correlation lengths for the fits in (a), with blue squares for the XY component and red triangles for the collinear antiferromagnetic component. (d) - (f) Same as (a) - (c) but for \NSCF. In panels (b), (c), (e), and (f), dashed curves are guides to the eye. }
	
\end{figure}
%%%%%%%%%%%%
% End Figure
%%%%%%%%%%%%

We now move on to the data collected at higher temperature. The same two-component model was refined at 5~K, 15~K, and 50~K, resulting in the fits displayed in Fig.~\ref{fig:both}(a). The refined scale factors for the XY and collinear AF components are displayed in Fig.~\ref{fig:both}(b) as blue squares and red triangles, respectively, and the refined correlation lengths are similarly displayed in Fig.~\ref{fig:both}(c). Inspection of the data shown in Fig.~\ref{fig:both}(a) reveals that the mPDF collected at 5~K is quite similar to that collected at 2~K, despite the fact that 5~K exceeds the spin freezing temperature \Tf\ = 2.4~K. This indicates that the short-range XY order persists just above \Tf\ in the form of correlated, dynamical fluctuations. At 15~K and 50~K, the mPDF patterns likewise show similar structure, but with progressively smaller amplitude and shorter extent in $r$. These observations are quantified by the refined the scale factor and correlation lengths in Fig.~\ref{fig:both}(b) and (c), respectively. For the XY component, the scale factor and correlation length both decrease steadily as the temperature is raised. The reduction of the scale factor is due to stronger uncorrelated fluctuations of the moments. The short-range collinear AF correlations remain nearly unchanged from 2~K to 15~K, but at 50~K, they have weakened considerably and are comparable in magnitude to the XY correlations. The data at 50~K are approaching the limits of our ability to distinguish the mPDF signal from the noise in the data. Nevertheless, these data show that magnetic correlations are still present at 50~K, significantly higher than the highest temperature previously studied (14~K).~\cite{ross;prb16} From the polarized neutron scattering data that will be presented shortly, we will see that in fact correlations persist up to 200~K. 

The results presented so far are fully consistent with the previous single-crystal neutron scattering measurements of \NCCF. Next, we consider the data collected from \NSCF, which we analyzed in precisely the same way. The best fit at 2~K using the two-component model is shown as the lowest set of curves in Fig.~\ref{fig:both}(d), with the black curve representing the fit and the gray curve the data. As with \NCCF, the XY model fits the data well for $r \gtrsim 7$~\AA, but the collinear AF component is necessary to fully capture the data at lower $r$, particularly the strong negative peak at the nearest-neighbor distance. Note that the sharp peak around 5~\AA\ not captured by the model is an artifact from the subtraction of the high-temperature fit residual. It is too sharp to be an mPDF peak; moreover, there are no pairs of Co atoms separated by that distance, so it cannot come from any magnetic correlations. The refined correlation length of the XY component for \NSCF\ is 8.7~$\pm$~0.4~\AA. This is again shorter than the previously reported length of 17.6~\AA. The temperature dependence of the mPDF data for \NSCF, seen for 15~K and 50~K in Fig.~\ref{fig:both}(d), is very similar to that of \NCCF. The main difference is that the negative peak at the nearest-neighbor distance at 50~K is stronger in \NSCF\ than \NCCF, suggesting that the dynamical collinear AF correlations are more robust in the former compound than in the latter. The refined scale factors and correlation lengths for the XY and collinear AF components are displayed in Fig.~\ref{fig:both}(e) and (f) as blue squares and red triangles. From this analysis, we draw the expected conclusion that the magnetic properties of \NCCF\ and \NSCF\ are essentially identical, in agreement with recent work on single-crystal samples~\cite{ross;prb17}.

\begin{figure}
	\includegraphics[width=80mm]{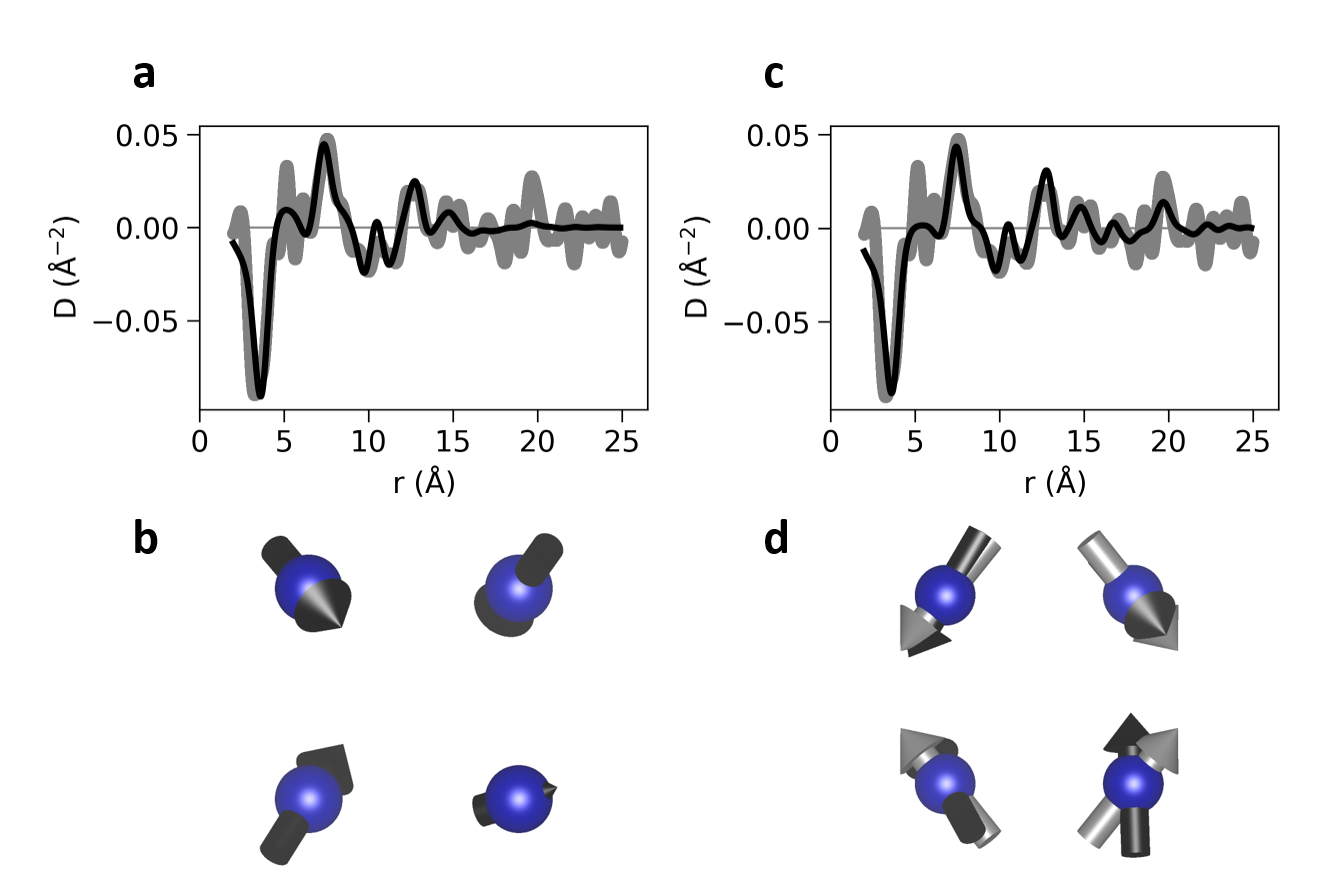}
	\caption{\label{fig:unconstrained} (Color online) (a) Magnetic PDF refinement of an unconstrained model where the direction of each magnetic moment in the unit cell is freely refined. The gray curve represents the data from \NSCF\ at 2~K and the black curve is the fit. (b) Refined magnetic moments projected onto a single tetrahedron and averaged. Collinear antiferromagnetic correlations cause the moment to be severely reduced on the lower-right site. (c) Same as (a), but for the model that includes the collinear antiferromagnetic component. (d) Same as (b), but for the fit in (c). The refined moments are shown as black arrows, and the most closely matching spin configuration drawn from the $\psi_2$ and $\psi_3$ basis states is overlaid with gray arrows, showing good agreement. }
	
\end{figure}

\textit{Robustness of the mPDF modeling.} A natural question arising from the preceding analysis is whether the mPDF data alone are sufficient to determine the local magnetic structure of \NCCF\ and \NSCF, or whether prior knowledge, such as that provided by the single-crystal neutron scattering measurements, is necessary to obtain a successful model. To investigate this, we performed a series of unconstrained refinements in which the known magnetic structure was not enforced. For the first of these unconstrained refinements, the orientations of all 16 magnetic moments in the unit cell were refined simultaneously and independently from each other, with the initial configuration consisting of completely randomly oriented moments. In addition to the 32 free parameters defining the 16 independent moment directions, we refined an exponential correlation length and a scale factor. The result of this fit for \NSCF\ at 2~K is shown in Fig.~\ref{fig:unconstrained}(a). The fit performs fairly well but is unable to capture all the features at higher $r$. Inspection of the configuration of refined magnetic moments reveals a tendency for the four tetrahedra in the unit cell to share a similar magnetic configuration, with one exception: on one of the tetrahedral sites, the moments on two of the tetrahedra are nearly parallel to each other but \textit{antiparallel} to the moments on the corresponding site of the other two tetrahedra, forming an approximately collinear AF arrangement. This is seen clearly in Fig.~\ref{fig:unconstrained}(b), where the four tetrahedra have been projected onto a single tetrahedron and averaged together. The average moment on one of the tetrahedral sites (lower-right position in the figure) is dramatically reduced due to this collinear AF arrangement. The other three average moments align roughly with a linear combination of $\psi_2$ and $\psi_3$, with the dot product of each spin with its match from the $\Gamma_5$ manifold coming to approximately 0.8.
%%%%%%%%%%%%%%%%%
% Begin Figure
%%%%%%%%%%%%%%%%%
%%%%%%%%%%%%
% End Figure
%%%%%%%%%%%%

This result, obtained without any prior assumptions aside from the absence of any ordering vector that changes the size of the unit cell, already hints at a two-component model consisting of XY-type correlations from the $\Gamma_5$ manifold and collinear AF correlations. Taking this a step further, we performed another refinement in which all 16 moments were freely refined, but we included a second component with the collinear AF arrangement enforced. We refined a scale factor and correlation length for each component in addition to the directions of the 16 unconstrained moments. This produces the fit shown in Fig.~\ref{fig:unconstrained}(c). In this case, the fit is excellent over the whole range. Furthermore, the unconstrained moment directions naturally converge to a configuration that is well aligned with a linear combination of $\psi_2$ and $\psi_3$, as shown by the comparison of the black arrows (refined moments averaged over all four tetrahedra) and the gray arrows (best match from the expected $\Gamma_5$ structure) in Fig.~\ref{fig:unconstrained}(d). The slightly greater misalignment of the lower-right moment is unlikely to have physical significance. Based on these free refinements, we can conclude that the mPDF data alone are sufficient to determine the SRO magnetic structure in \NSCF\ and \NCCF\ without any prior knowledge. The success of this approach for these materials portends well for its application to other frustrated magnetic systems in which little is known about the magnetic correlations.

As an additional check on the previous results, we performed an alternative fit to the \NSCF\ data at 2~K somewhat in the style of ``big-box'', atomistic modeling~\cite{egami;b;utbp12,keen;n15}. We populated a large sphere (radius 50~\AA) with magnetic moments at the Co positions determined from the structural PDF refinements. Initially, the moments on each tetrahedron were set to randomly generated linear combinations of $\psi_2$ and $\psi_3$, with no systematic inter-tetrahedral correlations. The mPDF calculated from such a configuration consists solely of a negative peak at the Co-Co nearest neighbor distance due to the antiferromagnetic intra-tetrahedral correlations of the $\psi_2$ and $\psi_3$ basis states; at larger $r$, the mPDF averages to zero.

Next, we programmed in short-range correlations by allowing each tetrahedron to duplicate the tetrahedron at the center of the sphere with a probability of $\exp(-r/\xi)$, where $r$ is the distance separating a given tetrahedron from the one at the center and $\xi$ is the chosen correlation length. We computed the mPDF from this configuration, then generated another configuration following the same protocol, averaged the new mPDF with the first, and continued this procedure for a total of 50,000 iterations. This was sufficient to ensure a very high level of convergence. The same collinear AF component used for the earlier mPDF modeling was then included, and scale factors were refined for both components, along with a correlation length for the collinear AF component, to produce the best fit to the experimental data.

This entire process was repeated using values of $\xi$ ranging from 5~\AA\ to 20~\AA\ in steps of 1.25~\AA. To estimate the magnetic correlation length, we calculated the goodness-of-fit parameter $R = \frac{\sum\left[ D_{\mathrm{obs}}-D_{\mathrm{calc}}\right]^2}{\sum\left[D_{\mathrm{obs}}\right]^2}$ for each value of $\xi$. A broad minimum occurs with its center around 10~$\pm$~2~\AA. This is in good agreement with the correlation length obtained from the simpler model, and the fits from the two approaches are indistinguishable from one another, confirming previous analysis.

We note that the above analyses represent the first quantitative modeling of a genuinely short-range-ordered magnetic ground state using the mPDF technique. Previously, this technique was used to investigate short-range correlations in the paramagnetic state of a conventional antiferromagnet~\cite{frand;prl16}, but not for a system lacking long-range order at all temperatures. Not only does this serve as a useful demonstration that the technique can be applied successfully to frustrated magnets and other systems with SRO magnetic ground states, but it also highlights the intuitive appeal of viewing short-range magnetic correlations directly in real space. By simple inspection of the data, one can see the clear decay of the correlations with $r$ and make a reasonable estimate of the correlation length without performing any analysis. In the present case, a simple model was then able to produce a quantitatively accurate fit to the mPDF and yield physically meaningful information about the magnetic structure. Moreover, the short-range magnetic structure could be obtained naturally by performing unconstrained refinements of the magnetic moment directions. The energy integration of the NOMAD instrument also provided us with sensitivity to the collinear AF component with no additional measurements(although without the ability to distinguish between fluctuating and static correlations), whereas comprehensive elastic and inelastic scattering measurements were required to identify this component in the earlier single-crystal work. 

\textit{Temperature-dependent polarized neutron scattering.}
%%%%%%%%%%%%%%%%%
% Begin Figure
%%%%%%%%%%%%%%%%%
\begin{figure}
	\includegraphics[width=65mm]{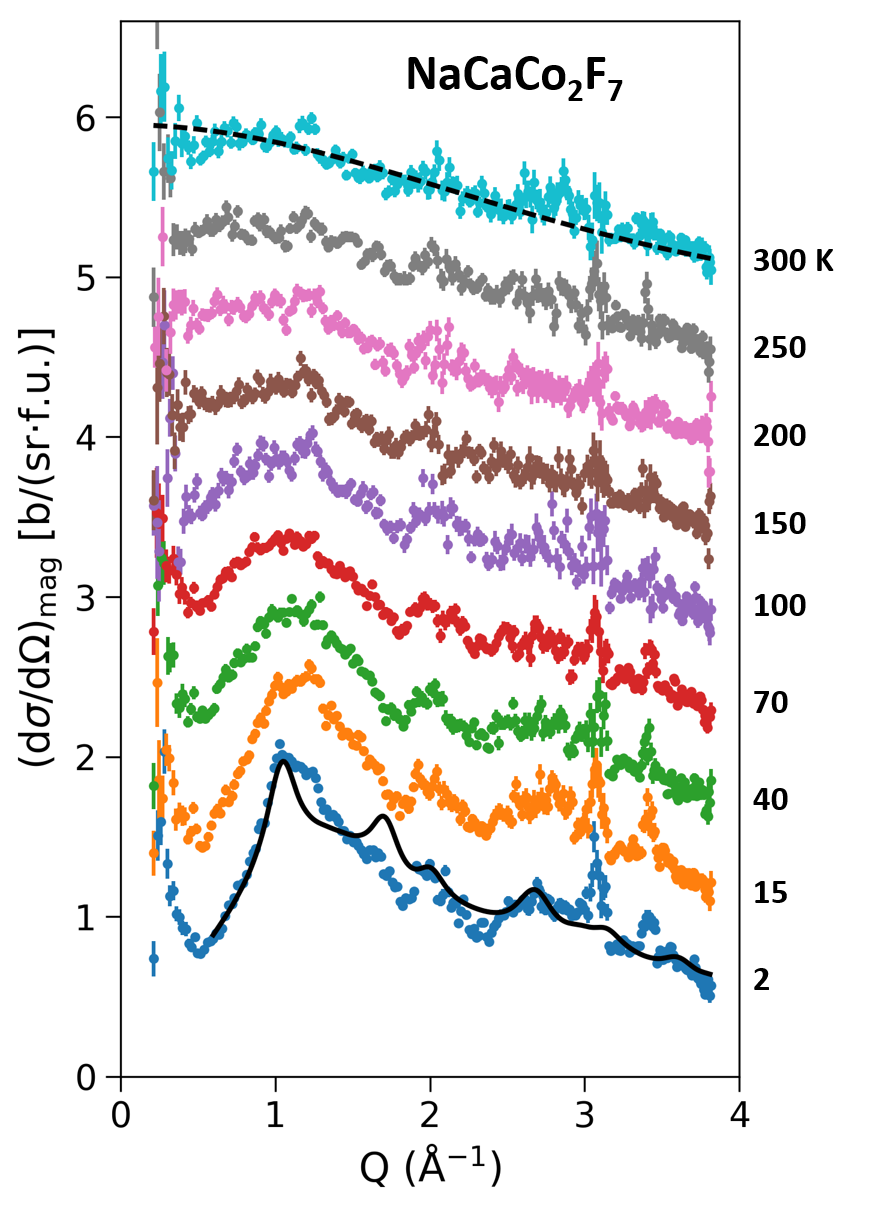}
	\caption{\label{fig:NCCF-D7} (Color online) (a) Magnetic scattering cross section at various temperatures obtained by polarized neutron scattering at the D7 instrument, offset vertically for clarity. Structure in the diffuse scattering is observed up to 200~K. The black dashed curve overlaid on the data collected at 300 K shows the square of the magnetic form factor. The black solid curve for 2~K represents a fit generated from the Fourier transform of the refined mPDF at that temperature.  The units on the vertical axis are barns per steradian per formula unit.}
	
\end{figure}
%%%%%%%%%%%%
% End Figure
%%%%%%%%%%%%
We now present the data obtained from polarized neutron scattering measurements conducted on the D7 instrument at ILL with the $xyz$ polarization method~\cite{schar;pssa93}. In this technique, 6 distinct scattering cross sections with different polarization and spin-flip conditions are collected at each temperature, and the magnetic scattering is isolated by taking appropriate linear combinations. Fig.~\ref{fig:NCCF-D7} displays the resulting magnetic scattering cross section of \NCCF\ collected at several temperatures ranging from 2~K to 300~K, offset vertically for clarity. First, we consider the data collected at 300~K (cyan circles). The magnetic scattering is entirely featureless and simply follows the square of the magnetic form factor (dashed black curve), indicating there are no appreciable magnetic correlations at 300~K. In contrast, the magnetic scattering at 2~K (dark blue circles) is quite structured, albeit still diffuse, as expected for short-range order. The greatest intensity is centered around 1~\AA$^{-1}$, consistent with the NOMAD data, and smaller humps of scattering can be seen at larger values of $Q$, most notably 2~\AA$^{-1}$ and 2.7~\AA$^{-1}$. At very low $Q$, a small increase in scattered intensity is observed, which is likely an artifact of the imperfect separation of magnetic and non-magnetic cross sections due to the presence of inelastic magnetic scattering~\cite{ehler;rsi13}. The temperature-independent features around 2.0~\AA$^{-1}$, 3.0~\AA$^{-1}$, and 3.5~\AA$^{-1}$ are likewise caused by imperfect separation of nuclear Bragg peaks near those positions.

These data in momentum space invite comparison to the mPDF data obtained previously. To do so, we Fourier transform the refined mPDF back into momentum space. The result of this is shown as the black curve overlaid on the dark blue data points for 2~K in Fig.~\ref{fig:NCCF-D7}, showing good agreement. The effects of inelastic scattering are evidently not overly problematic in the present case due to the low energy scale of the inelastic magnetic scattering in \NCCF. We note that the black curve in Fig.~\ref{fig:NCCF-D7} was fit to the scattering data by refining a scale factor for the Fourier transformed mPDF and including the magnetic form factor, allowed to vary slightly from the empirical result expected for Co$^{2+}$ with a full orbital contribution~\cite{itable;volc95}. The consistency between the refined mPDF and the polarized neutron scattering data is further evidence that the model of XY and collinear AF correlations is correct and that the mPDF analysis presented previously is reliable. The success of the mPDF measurements, which are experimentally quite simple compared to the polarized neutron measurements, is evidence of the power of the mPDF technique even when the signal is quite small.

Now we examine the temperature dependence of the magnetic scattering. With increasing temperature, the structured features gradually lose intensity and become increasingly poorly defined. At 70~K, only a broad hump around 1~\AA$^{-1}$ remains. However, this hump is remarkably robust, remaining visible even in the data collected at 200~K. Only for 250~K and 300~K is it entirely absent. This demonstrates that magnetic correlations persist up to at least 200~K in \NCCF, approximately two orders of magnitude larger than the spin freezing temperature. We suggest that this persistent high-temperature diffuse scattering is primarily due to the short-range collinear AF fluctuations, since the strongest feature in the mPDF data at the highest measured temperature was the nearest-neighbor AF peak. The presence of these correlations at such high temperatures demonstrates that the intrinsic magnetic interactions in \NCCF\ are quite strong and is consistent with the high Curie-Weiss temperature $|\Theta_{\mathrm{CW}}|\approx 140$~K; that the spin freezing temperature is so low highlights the dramatic effects of geometric frustration.

\section{Conclusion}
We have reported temperature-dependent neutron total scattering and polarized neutron scattering measurements of \NCCF\ and \NSCF. The polarized neutron data revealed magnetic correlations in \NCCF\ that persist up to 200~K, which we identified as primarily near-neighbor AF correlations based on mPDF analysis of the neutron total scattering data. These findings demonstrate that the magnetic interactions in these new fluoride pyrochlore systems are strong, supporting the notion that they offer the valuable opportunity to study frustration-related effects at significantly higher temperatures than is possible for oxide pyrochlores. Additionally, through mPDF modeling of the low-temperature neutron total scattering data, we confirmed that the basic nature of the magnetic ground state of \NCCF\ and \NSCF\ consists of short-range XY-type correlations with a characteristic length scale of $\sim$~10~\AA\ accompanied by collinear AF fluctuations restricted primarily to near-neighbor moments~\cite{ross;prb16,ross;prb17}.

This work has also advanced the mPDF technique. We have demonstrated that this method can be applied successfully to geometrically frustrated systems possessing SRO ground states, allowing an analysis of local magnetic correlations directly in real space. In the present case, unconstrained fits were able to determine the short-range magnetic structure directly with a high degree of accuracy. Because mPDF measurements are performed on powder samples, this technique can be used even when single-crystal specimens are unavailable. We have also presented a procedure for extracting small mPDF signals from non-ideal PDF data in which backgrounds and imperfections in the atomic PDF refinement are comparable to the magnitude of the mPDF. We expect this approach to be useful for future mPDF experiments. Finally, this work represents the first direct comparison between mPDF analysis and polarized neutron scattering measurements, providing further experimental verification of the mPDF technique.

\textbf{Acknowledgements}

We thank Matt Tucker and Kate Page for assistance at ORNL. Work at Lawrence Berkeley National Laboratory and University of California, Berkeley was supported by the Office of Science, Office of Basic Energy Sciences, Materials Sciences and Engineering Division, of the US Department of Energy (US DOE) under Contract No. DE-AC02-05-CH11231 within the Quantum Materials Program (KC2202) and the Office of Basic Energy Sciences, US DOE, Grant No. DE-AC03-76SF008. SJLB's contributions to the mPDF data interpretation and writing were supported by the US DOE under contract No. DE-SC00112704. The crystal growth at Princeton University was supported by the US DOE Office of Basic Energy Sciences, Grant No. DE-FG02-08ER46544. Use of the Spallation Neutron Source, Oak Ridge National Laboratory, was sponsored by the Scientific User Facilities Division, Office of Basic Energy Science, U.S. DOE. We thank the Institut Laue-Langevin for the use of its neutron instrumentation.

\end{document}